\documentclass[letterpaper, 10 pt, conference]{ieeeconf}  
\IEEEoverridecommandlockouts
\usepackage[utf8]{inputenc}
\usepackage[T1]{fontenc}
\usepackage{amsmath}
\usepackage{amsfonts}
\usepackage{graphicx}
\usepackage{xcolor}
\usepackage{figsize}
\usepackage{upgreek}
\usepackage{amssymb}
\usepackage{algorithmic}
\usepackage[linesnumbered,ruled,vlined]{algorithm2e}
\usepackage{bm}
\usepackage{xargs}
\usepackage{soul}
\usepackage{cancel}
\usepackage{mathtools}
\usepackage{verbatim}
\usepackage{graphics}
\usepackage{epsfig} 
\usepackage{xfrac}
\usepackage{tikz}
\usepackage{accents}
\usepackage[normalem]{ulem}
\usepackage{booktabs}
\usepackage{siunitx}
\graphicspath{{figures/}}

\newcommand{\norm}[1]{\left\lVert#1\right\rVert}

\newcommand{\x}{{\mathbf{x}}}
\newcommand{\y}{{\mathbf{y}}}
\renewcommand{\u}{{\mathbf{u}}}

\newcommand{\bx}{{\x}}
\newcommand{\bu}{{\u}}

\newcommand{\bg}{{{b_g}_{n|k}}}
\newcommand{\DDelta}{{\mathbf{\Delta}}}

\newcommandx{\xb}[2][1=n,2=k]{\x_{#1|#2}}

\newcommandx{\zb}[2][1=n,2=k]{\bar\z_{#1|#2}}
\newcommandx{\ub}[2][1=n,2=k]{\u_{#1|#2}}
\newcommandx{\yb}[2][1=n,2=k]{\bar\y_{#1|#2}}
\newcommandx{\vb}[2][1=n,2=k]{\vv_{#1|#2}}
\newcommandx{\rbx}[2][1=n,2=k]{\r_{#1|#2}^{\x}}
\newcommandx{\rbu}[2][1=n,2=k]{\r_{#1|#2}^{\u}}
\newcommandx{\hb}[3][1=n,2=k,3={}]{h_{#1}^{#3}}
\newcommandx{\gb}[3][1=n,2=k,3={}]{g_{#1|#2}^{#3}}
\newcommandx{\gbar}[3][1=n,2=k,3={}]{\bar{g}_{#1|#2}^{#3}}
\newcommandx{\xT}[2][1=n,2=k]{\mathcal{X}_{#1|#2}}

\newtheorem{Theorem}{Theorem}

\newtheorem{Proposition}{Proposition}

\newtheorem{Assumption}{Assumption}
\newtheorem{Definition}{Definition}

\title{\LARGE \bf
Priority-driven Constraints Softening in Safe MPC for Perturbed Systems}

\author{Ying Shuai Quan$^{1}$, Mohammad Jeddi$^{2,3}$, Francesco Prignoli$^{2}$ and Paolo Falcone$^{1,2}$
\thanks{This work is funded by the Fordonsstrategisk
Forskning och Innovation program of Vinnova under the grant 2018-05005.
		}
        \thanks{Project co-funded by the European Union – Next Generation Eu - under the National Recovery and Resilience Plan (NRRP), Mission 4 Component 1 Investment 4.1 - Decree No 118 of Italian Ministry of University and Research - Concession Decree No. 2333 of the Italian Ministry of University and Research, Project code D93C23000450005, within the Italian National Program PhD Programme in Autonomous Systems (DAuSy).
		}
\thanks{$^{1}$ Ying Shuai Quan and Paolo Falcone are with the Mechatronics group at the Department of Electrical Engineering, Chalmers University of Technology, Gothenburg, Sweden {\tt\footnotesize \{quany,paolo.falcone\}@chalmers.se }}
\thanks{$^{2}$ Mohammad Jeddi, Francesco Prignoli, and Paolo Falcone are with the Dipartimento di Ingegneria ``Enzo Ferrari'' Universit\`a di Modena e Reggio Emilia, Italy {\tt\footnotesize \{mohammad.jeddi, francesco.prignoli, falcone\}@unimore.it }}
\thanks{$^{3}$ Mohammad Jeddi is also with  the Dept. of Electrical and Information Engineering, Italian National Ph.D. DAUSY, Polytechnic of Bari, Italy {\tt\footnotesize \ m.jeddi@phd.poliba.it}}
}

\begin{document}
\maketitle
\thispagestyle{empty}
\pagestyle{empty}
\begin{abstract}
This paper presents a \emph{safe} model predictive control (SMPC) framework designed to ensure the satisfaction of \emph{hard} constraints, for systems perturbed by an external disturbance. Such safety guarantees are ensured, despite the disturbance, by online softening a subset of adjustable constraints defined by the designer. The selection of the constraints to be softened is made online based on a predefined priority assigned to each adjustable constraint. The design of a learning-based algorithm enables real-time computation while preserving the original safety properties.

Simulations results, obtained from an automated driving application, show that the proposed approach provides guarantees of collision-avoidance \emph{hard} constraints despite the \emph{unpredicted} behaviors of the surrounding environment.
\end{abstract}
\section{Introduction}
Model Predictive Control (MPC) is widely used in safety-critical systems by incorporating preview information and constraints. However, in practice, not all constraints are known in the design stage. The presence of a-priori unknown constraints, which may change over time, can lead to the loss of feasibility. To address this issue, safe model predictive control (SMPC) frameworks have been developed to ensure the satisfaction of a-priori unknown constraints at all times, provided that a set of assumptions holds~\cite{batkovic2022safe,batkovic2023experimental}. However, deploying these frameworks in real-world scenarios remains challenging, as unforeseen environmental factors can still violate the required assumptions, potentially leading to infeasibility.

To address the challenges posed by varying environments, several studies have proposed MPC frameworks that incorporate time-varying constraints~\cite{liu2019recursive}. However, these approaches often rely on predicting the maximal environmental variation, which can lead to overly conservative control strategies. Moreover, handling time-varying constraints increases the complexity of controller design, particularly due to the necessity for online computation of the time-varying terminal set~\cite{manrique2014mpc}. 
An alternative approach introduces slack variables to soft constraints, improving adaptability to environmental changes~\cite{rakovic2021model}. This method also eliminates the need for explicit computation of a time-varying terminal set by incorporating terminal dynamics for the slack variable. 
%
  {
In addition, the design of an exact penalty function ensures that, whenever a feasible solution to the original hard-constrained problem exists, the optimal solution of the slacked problem coincides with that of the original problem \cite{kerrigan_soft_2000,borrelli_predictive_2017}. However, determining which soft constraints are ultimately relaxed depends on the penalty weights assigned to each slack variable—a choice that becomes increasingly challenging as the number of slack variables grows. In \cite{kerrigan_designing_2002}, the authors address this issue by proposing a MPC design with prioritized constraints, formulated as a multi-objective optimization problem and solved through a sequence of single-objective MPCs. While this approach provides a structured way to prioritize constraints, it can be computationally expensive for online implementation.}

Motivated by these challenges, we propose a priority-driven constraint softening approach in SMPC, that dynamically adjusts a set of adjustable constraints in real time to ensure the satisfaction of hard constraints under unforeseen external disturbances.
Our approach leverages a learning-based algorithm to efficiently approximate the relationship between disturbances and constraint adaptations. By structuring constraint relaxations based on predefined priorities, we ensure safe and feasible control while improving computational efficiency. We validate our method in an autonomous driving scenario, demonstrating its ability to handle unforeseen constraints and maintain safety, making it a promising solution for real-world deployment.

\subsection{Notation}\label{sec:notation}
Consider a discrete-time linear system described by
	\begin{equation}\label{eq:sys}
	\x_{k+1}=A\x_k+B\u_k,		
	\end{equation}
	where $\x_k\in\mathbb{R}^{n_x}$ and $\u_k\in\mathbb{R}^{n_u}$ are the state and input vectors at time $k$, respectively. 
	The system is subject to two groups of convex state and input constraints: \emph{a-priori known} constraints $h(\x,\u)\leq{}0$, given by
	\begin{equation}
			h(\x_{n|k},\u_{n|k}) = C_h\x_{n|k} +D_h\u_{n|k} - {\mathbf{b}_h},
	\end{equation}
	and \emph{a-priori unknown}   {\emph{hard}} constraints 
    $ {g(\x,\u, b_g)\leq{}0}$ in the form of
     {
	\begin{equation}
			g(\x_{n|k},\u_{n|k},{b_g}_{n|k}) = C_g\x_{n|k} +D_g\u_{n|k} - {b_g}_{n|k},
	\end{equation}
      {where~${b_g}_{n|k}$ is an external disturbance. While the constraints~$h(\cdot,\cdot)$ are defined at the problem formulation stage}, the constraints $g(\cdot,\cdot,\cdot)$ are constructed online, as the system evolves, and are driven by an external input. For example, in the context of mobile autonomous systems, $g(\cdot,\cdot,\cdot)$ 
      {arise from collision-avoidance} constraints 
    imposed by the interactions with the environment and the disturbance~${b_g}_{n|k}$ is the predicted motion of surrounding moving obstacles.
    }
	  {While the proposed formulation works for an arbitrary number of known and unknown constraints, for simplicity, we here consider multiple a priori-known constraints, such that $h(\x,\u)\in \mathbb R^{n_h}$, and one single a priori-unknown constrain, such that $g(\x,\u,b_g)\in \mathbb R^1$.}
	The predicted state, input and 
      {disturbance} at time $n$ given the information available at time $k$ are denoted by $\x_{n|k}$, $\u_{n|k}$, and $ {{b_g}_{n|k}}$ respectively. 
	We use the notation $\mathbb{I}_{a}^{b} := \{a,a+1,...,b\}$ to denote a set of integers.
	We denote the Euclidean norm as $\norm{x} = \sqrt{x^T x}$ whereas the Euclidean norm w.r.t. $P=P^T\succ 0$ is denoted by $\| x \|^2_P=x^T P x$.
     {We define $\x^\u_+$ as the next state under control input $\u$: $\x^\u:=A\x+B\u$.}
   
\section{Safe MPC}
We define safety, which is the primary objective of our controller:
	\begin{Definition}[Safety]\label{def:safe}
		A controller $\u_k = \kappa(\x_k)$ is safe for the system (\ref{eq:sys}) in a set $\mathcal{S}\subseteq\mathbb{R}^{n_x}$ if $\forall \, \x\in\mathcal{S}$, the control inputs $\mathbf{U}=\{\kappa(\x_0),...,\kappa(\x_\infty)\}$ and the corresponding state trajectories $\mathbf{X}=\{\x_0,\x_1,...,\x_\infty\}$ are such that $h(\x_k,\u_k)\leq{}0$ and~ {$g(\x_k,\u_k,{b_g}_k)\leq{}0$}, $\forall \, k \geq 0$.
	\end{Definition}

Based on the Safety Definition \ref{def:safe}, we formulate the following SMPC control problem. 
    \begin{subequations}\label{eq:nmpc}
		\begin{align}
		\min_{\substack{\bu}}& \sum_{n=k}^{k+N-1}
		q_\mathbf{r}(\xb,\ub)+p_\mathbf{r}(\xb[k+N])\nonumber\\
		\text{s.t.}\ &\xb[k][k] = \x_{k},&\label{eq:nmpcState} \\
		&\xb[n+1] = A\xb+B\ub, &\hspace{-3em}n\in \mathbb{I}_k^{k+M-1} \label{eq:nmpcDynamics}\\
		&h(\xb,\ub) \leq{} 0, & \hspace{-3em}n\in \mathbb{I}_k^{k+M-1} \label{eq:nmpcInequality_known}\\
		& {g(\xb,\ub,\bg)} \leq{} 0, & \hspace{-3em}n\in \mathbb{I}_k^{k+M-1} \label{eq:nmpcInequality_unknown}\\
		&\xb[k+n] \in\mathcal{X}_\mathbf{r}^\mathrm{s},&\hspace{-3em}n\in \mathbb{I}_{k+N}^{k+M-1} \label{eq:nmpcTerminal}\\
		& {\xb[k+M]\in\mathcal{X}_\mathrm{safe}},\hspace{-10em}&\label{eq:nmpcTerminalSafe}
		\end{align}
	\end{subequations}
    where $k$ is the current  time, $N$ is the prediction horizon, while $M \geq N$ is an extended  prediction horizon. 
	The stage and terminal cost functions $q_\mathbf{r}$ and $p_\mathbf{r}$ penalize the deviations from the reference trajectory $\mathbf{r} = (\mathbf{r}^\x, \mathbf{r}^\u)$. 
	The constraint \eqref{eq:nmpcState} ensures that predictions of the systems dynamics in \eqref{eq:nmpcDynamics} start from the current system state~$\x_{k}$. 
	The constraints \eqref{eq:nmpcTerminal} and \eqref{eq:nmpcTerminalSafe} define a stabilizing and a safe set, respectively. It can be shown that the system
    \begin{equation}\label{eq:sys_cl}
	\x_{k+1}=A\x_k+B\kappa(\x_k),
	\end{equation}
    with~$\kappa(\x_k)=\ub^*$ and~$\{\ub^*\}_{n=k}^{k+M-1}$ solution of~\eqref{eq:nmpc}, is safe according to Definition~\ref{def:safe}, if the following assumptions hold.

    \begin{Assumption}\label{a:cont}
			The pair $(A,B)$ is stabilizable. 
             {The stage cost $q_\mathbf{r}:\mathbb{R}^{n_x}\times \mathbb{R}^{n_u} \rightarrow\mathbb{R}_{\geq{}0}$, and terminal cost $p_\mathbf{r}:\mathbb{R}^{n_x}\rightarrow\mathbb{R}_{\geq{}0}$, are continuous at the origin and satisfy 
            $q_\mathbf{r}({\mathbf{r}^{\x}_k},\mathbf{r}^{\u}_k)=0$, and $p_\mathbf{r}({\mathbf{r}^{\x}_k})=0$.}
			 Additionally, $q_\mathbf{r}(\bx_k,\bu_k)\geq{}\alpha_1(\|\bx_k-{\mathbf{r}^{\x}_k}\|)$ for all feasible $\x_k$, $\u_k$, and  $p_\mathbf{r}(\bx_N)\leq\alpha_2(\|\bx_N-\mathbf{r}^{\x}_N\|)$, where $\alpha_1$ and $\alpha_2$ are $\mathcal{K}_\infty$-functions.
	\end{Assumption}
    
    \begin{Assumption} \label{a:rec_ref}
			The reference is feasible for the system dynamics, i.e., $\mathbf{r}^\x_{k+1}=A\mathbf{r}^\x_k+B\mathbf{r}^\u_k$, and satisfies the known constraints \eqref{eq:nmpcInequality_known}, i.e., 
             {$h(\mathbf{r}^\x_k,\mathbf{r}^\u_k) \leq{} 0$, for all $k\in\mathbb{I}_0^\infty$.}
	\end{Assumption}
	 {\begin{Assumption} \label{a:terminal}
		There exists a parametric stabilizing terminal set $\mathcal{X}^\mathrm{s}_\mathbf{r}$ and a terminal control law $\kappa^\mathrm{s}_\mathbf{r}(\mathbf{x})$ yielding $p_\mathbf{r}(\x^{\kappa^\mathrm{s}_\mathbf{r}(\mathbf{x})}) - p_\mathbf{r}(\x) \leq{} - q_\mathbf{r}(\x,\kappa^\mathrm{s}_\mathbf{r}(\x)$,  $\x^{\kappa^\mathrm{s}_\mathbf{r}(\mathbf{x})}\in\mathcal{X}^\mathrm{s}_\mathbf{r}$, and 
        $h(\x,\kappa^\mathrm{s}_\mathbf{r}(\x)) \leq{} 0$.
	\end{Assumption}}

	Assumptions \ref{a:cont}-\ref{a:terminal} are commonly used in standard MPC to enforce stability, as noted in \cite{rawlings2017model}. 
	To additionally ensure safety, we introduce the following assumptions:
	\begin{Assumption} \label{a:unknown_constraints}
    The a-priori unknown constraint function satisfies
     {
        \begin{align}
			g(\xb,\ub,{b_g}_{n|k+1}) &\leq g(\xb,\ub,\bg),
            \label{eq:assump_unrelaxed}
		\end{align}
        }
    for all $n\geq k$.
	\end{Assumption}
	Assumption~\ref{a:unknown_constraints} requires that the a-priori unknown constraints $g(\x_k,\u_k,{b_g}_k)\leq{}0$ are \emph{consistent}, that is, they do not become ``more restrictive'' as the system~\eqref{eq:sys} evolves.   {In the context of autonomous mobile robots, Assumption~\ref{a:unknown_constraints} requires that, at time~$k+1$, a moving obstacle does not get ``much'' closer to the robot than what has been predicted at time~$k$.}
	Finally, the following assumption requires the existence of a safe set, which helps ensuring that the controller's recursive feasibility holds. 
     {\begin{Assumption}\label{a:safe}
		There exists a robust invariant set denoted $\mathcal{X}_\mathrm{safe}\subseteq\mathcal{X}_\mathbf{r}^\mathrm{s}$ such that for all $\x\in \mathcal{X}_\mathrm{safe}$ there exists a safe control set $\mathcal{U}_\mathrm{safe}\subseteq\mathbb{R}^{n_u}$ entailing that  $\x^{\u_\mathrm{safe}}\in\mathcal{X}_\mathrm{safe}$, and $h(\x,\u_\mathrm{safe})\leq{}0$, for all $\u_\mathrm{safe}\in\mathcal{U}_\mathrm{safe}$. 
		Moreover, by construction $g(\x,\u_\mathrm{safe},\bg) \leq 0$ for all $\x\in \mathcal{X}_\mathrm{safe}$ and $\u_\mathrm{safe}\in\mathcal{U}_\mathrm{safe}$.
	\end{Assumption}}

	Based on the assumptions \ref{a:cont}-\ref{a:safe}, the following result guarantees the safety of the controller in (\ref{eq:nmpc}).
	\begin{Proposition}
	 \cite{batkovic2022safe}
		\label{lemma:1}
		Suppose that Assumptions \ref{a:cont}, \ref{a:rec_ref}, \ref{a:terminal}, \ref{a:unknown_constraints}, and \ref{a:safe} hold, and that Problem (\ref{eq:nmpc}) is feasible for the initial state $\x_k$. 
		Then, system (\ref{eq:sys}) in a closed loop with the solution of (\ref{eq:nmpc}) applied in receding horizon is safe (recursively feasible) at all times.
		
		\begin{proof}
			 Readers are referred to \cite{batkovic2022safe}.
		\end{proof}
	\end{Proposition}

\section{{Priority-driven softened Safe MPC}}
Assumptions \ref{a:cont}-\ref{a:safe} enable strong safety guarantees \cite{batkovic2022safe}. However, Assumption \ref{a:unknown_constraints} can be overly restrictive   {and often violated in practice}. 
  {In this section, we design a learning-based, 
disturbance-responsive SMPC built upon a constraint relaxation mechanism that reacts to the disturbance~$b_g$, which causes the violation of Assumption \ref{a:unknown_constraints}. The extent of the violation of the Assumption~\ref{a:unknown_constraints} is described by~${\Delta_g}_{n|k+1} \in \mathbb{R}^+$, which is defined as}
\begin{equation}
{\Delta_g}_{n|k+1} = {b_g}_{n|k} - {b_g}_{n|k+1},
\label{eq:g_update}
\end{equation}
 {
and used to drive the selective relaxation of the a-priori known constraints~$h(\xb,\ub)\leq{} 0$ to ensure the persistent satisfaction of the hard constraints~$g(\xb,\ub,\bg) \leq{} 0$.}   {As shown in section~\ref{sec:unc_Resp_SMPC}, the proposed constraints relaxation mechanism can be computationally demanding.} Hence, we develop a learning-based algorithm that approximately determines the  necessary constraint relaxation at the cost of limited computational overhead. Finally, we present a recursive feasibility result under the proposed constraints relaxation mechanism.

\subsection{Uncertainty-Responsive SMPC}\label{sec:unc_Resp_SMPC}
Given the current a-priori unknown constraint $g(\xb,\ub,  {\bg})$, we solve the following optimization problem:
 {
\begin{subequations}\label{eq:slack_opt}
		\begin{align}
		\min_{\substack{\bu,\DDelta_h}}& 
		\sum_{n=k}^{k+N-1} q_h({\DDelta_h}_{n|k}) + p_h({\DDelta_h}_{k+N|k})\nonumber\\
		\text{s.t.}\ & (\ref{eq:nmpcState}),(\ref{eq:nmpcDynamics}), (\ref{eq:nmpcInequality_unknown}),(\ref{eq:nmpcTerminalSafe})& \\
        &0 \leq {\DDelta_h}_{n|k}, &\hspace{-3em}n\in \mathbb{I}_k^{k+M-1} \\
		&h(\xb,\ub) \leq{} E{\DDelta_h}_{n|k},  &\hspace{-3em}n\in \mathbb{I}_k^{k+M-1} \label{eq:relaxed_h}\\
		&(\xb, {\DDelta_h}_{n|k}) \in {{\mathcal{Z}}}^\mathrm{s}_\mathbf{r},&\hspace{-3em}n\in \mathbb{I}_{k+N}^{k+M-1} 
		\end{align}
	\end{subequations}
    }
\noindent where ${\DDelta_h}_{n|k}$ and $\u_{n|k}$ are decision variables. 
 {Assuming that $\DDelta_h \in \mathbb{R}^{n_\Delta}$,
$E\in \mathbb{R}^{n_h\times n_\Delta}$ is a user-defined  matrix that selects those constraints that can be relaxed, where $E_{ij} = 1$ if the $i$-th row can be relaxed with $j$-th element of $\DDelta_h$, and $E_{ij} = 0$ otherwise.   {Furthermore,~$\sum_{j=1}^{n_\Delta}~E_{ij}=1$ holds~$\forall~i\in\{1,\ldots,n_h\}$.}
  {We impose the terminal dynamics ${\DDelta_h}^+ = M_h\DDelta_h$, with }
  {$M_h\in \mathbb{R}^{{n_\Delta}\times {n_\Delta}}$ such that~$\DDelta_h \in \mathbb{R}^{{n_\Delta}}_+ \rightarrow M_h\DDelta_h\in \mathbb{R}^{{n_\Delta}}_+$ and~$\max\|\lambda(M_h)\|<1$}. Then by solving $M^\top_h P_h M_h - P_h \preceq -Q_h$ with $Q_h \in \mathbb{R}^{{n_\Delta}\times {n_\Delta}}, Q_h=Q^\top_h\succ 0$, the stage and terminal cost functions are defined as  $q_h:=\norm{{\DDelta_h}_{n|k}}_{Q_h}$ and $p_h:=\norm{{\DDelta_h}_{k+N|k}}_{P_h}$.
The introduction of the terminal dynamics for $\DDelta_h$ is intended to induce desirable terminal and asymptotic behaviors   {of the relaxation terms~${\DDelta_h}_{n|k}$.}}
For the extended terminal set ${{\mathcal{Z}}}^\mathrm{s}_\mathbf{r}$, we make the following assumption based on Assumption \ref{a:terminal}:
\begin{Assumption} \label{a:terminal_relaxed}
	There exists a terminal set ${{\mathcal{Z}}}^\mathrm{s}_\mathbf{r}$ 
    such that with the terminal control law $\kappa^\mathrm{s}_\mathbf{r}(\mathbf{x})$, $\forall (\x, {\DDelta_h})  \in {{\mathcal{Z}}}^\mathrm{s}_\mathbf{r} \Rightarrow (\x^{\kappa^\mathrm{s}_\mathbf{r}(\mathbf{x})}, M_h{\DDelta_h})\in {\mathcal{Z}}^\mathrm{s}_\mathbf{r}$, 
	and $h(\x,\kappa^\mathrm{s}_\mathbf{r}(\x),M_h{\DDelta_h}) \leq{} 0$.
\end{Assumption}

Denoting the result of (\ref{eq:slack_opt}) as ${\DDelta^\star_h}_{n|k}$ and substituting it into the SMPC framework (\ref{eq:nmpc}), we formulate the following uncertainty-responsive SMPC control problem:
 {\begin{subequations}\label{eq:unmpc}
		\begin{align}
		\min_{\substack{\bu}}& \sum_{n=k}^{k+N-1}
		q_\mathbf{r}(\xb,\ub)+p_\mathbf{r}(\xb[k+N])\nonumber\\
		\text{s.t.}\ &(\ref{eq:nmpcState}), (\ref{eq:nmpcDynamics}), 
        (\ref{eq:nmpcInequality_unknown}), (\ref{eq:nmpcTerminalSafe})& \\
		&h(\xb,\ub) \leq{} E{\DDelta^\star_h}_{n|k}, & \hspace{-3em}n\in \mathbb{I}_k^{k+M-1} \\
		&(\xb. {\DDelta^\star_h}_{n|k}) \in {{\mathcal{Z}}}^\mathrm{s}_\mathbf{r},&\hspace{-3em}n\in \mathbb{I}_{k+N}^{k+M-1} 
		\end{align}
	\end{subequations}}

  {The proposed disturbance-responsive SMPC framework requires solving the problems~\eqref{eq:slack_opt} and~\eqref{eq:unmpc}, which may not be possible in real-time. Hence, we next propose to replace the solution of~\eqref{eq:slack_opt} with a learning-based approximation.}

\subsection{ {Learning Uncertainty-Responsive SMPC}}
  {In this section we formulate the design and training problem of a Neural Network (NN) that approximates the solution of problem~\eqref{eq:slack_opt}.  {The input and output signals of the NN are defined as
\begin{equation}
	\boldsymbol \theta_k = 
	\begin{bmatrix}
		\begin{bmatrix}
		{b_g}_{n|k}
	\end{bmatrix}_{n\in \mathbb{I}_k^{k+M-1}}\\
		\x_k
	\end{bmatrix}, 
	\boldsymbol \eta_k = 
    \begin{bmatrix}
		{\DDelta_h}_{n|k}
	\end{bmatrix}_{n\in \mathbb{I}_k^{k+M-1}}.
    \label{eq:inout_NN}
\end{equation}}}

We introduce the following assumption
\begin{Assumption}\label{a:fea_set}
	There exists a nonempty compact set 
	${\Theta} = \{ 
		\boldsymbol{\theta}: (\ref{eq:slack_opt}) \ \text{is feasible}
	 \}.$
\end{Assumption}

\subsubsection{NN with slope-restricted nonlinearity}
We introduce an $l$-layer NN described by the following recursive equations:
\begin{equation}
\begin{split}
&\boldsymbol{s}^0 = \boldsymbol{\theta}, \boldsymbol{s}^{i+1} =
\phi_i(W^i \boldsymbol{s}^i + b^i), i\in \mathbb{I}^{l-1}_0,\\
&\boldsymbol{\eta} = f(\boldsymbol{\theta}) = W^{l} \boldsymbol{s}^{l} + b^{l},
\end{split}
\label{eq:NN}
\end{equation}
where $W^i \in \mathbb{R}^{n_{i+1} \times n_i}$, $b^i \in \mathbb{R}^{n_{i+1}}$, and $n_0, \ldots, n_{l+1}$ are the dimension of the input, the neurons in the hidden layers, and the output. The function $\phi_i$ is the vector of activation functions at each layer of the form 
$\phi_i(z^{i}) = 
\begin{bmatrix}
	\psi(z^{i}_1) \cdots \psi(z^{i}_n)
\end{bmatrix}^\top
$ 
with continuous nonlinear activation functions $\psi: \mathbb{R} \rightarrow \mathbb{R}$, with slope-restricted nonlinearity:
\begin{equation}
\alpha \leq
\frac{\psi(x)-\psi(y)}{x-y}
\leq \beta ~~~
\forall x,y \in \mathbb{R}.
\label{eq:slope}
\end{equation}

With a diagonal weighting matrix
$
T \in \mathcal{D}_n : =
\{
T = \sum_{i=1}^{n} \lambda_{ii} e_i e^T_i,
\lambda_{ii} \geq 0
\}
\label{eq:Tn}
$, a quadratic constraint can be used to characterize all the stacked activation functions:
\begin{equation}
\begin{bmatrix}
\tilde{x} - \tilde{y}\\
\phi(\tilde{x}) - \phi(\tilde{y})
\end{bmatrix}^T
F(T)
\begin{bmatrix}
\tilde{x} - \tilde{y}\\
\phi(\tilde{x}) - \phi(\tilde{y})
\end{bmatrix}
 \geq 0
\label{eq:layerslope}
\end{equation}
with
$
F(T)=
\begin{bmatrix}
-2\alpha\beta T & (\alpha+\beta)T\\
(\alpha+\beta)T & -2T
\end{bmatrix}
$
for all $T \in \mathcal{D}_n, \tilde{x}, \tilde{y} \in \mathbb{R}^n$ with $n=\sum_{i=1}^{l}n_i$.
Here, the inputs to all neurons in the $l$-layer NN are stacked up into one vector
$\tilde{x}, \tilde{y}$, respectively,
and the activation function for all layers
$
\phi :
\mathbb{R}^n
\rightarrow
\mathbb{R}^n,
\phi(\tilde{x}) =
\begin{bmatrix}
\phi_1(\tilde{x}^1)^T&
\cdots&
\phi_l(\tilde{x}^{l})^T
\end{bmatrix}
$
is applied to the concatenated vector.

\subsubsection{Lipschitz continuity analysis}
Firstly, we introduce the concept of Lipschitz continuity.
\begin{Definition}
	A function $f: \mathbb{R}^n \rightarrow \mathbb{R}^m$ is globally Lipschitz continuous if there exists an $L \geq 0$ such that:
\begin{equation}
\| f(x)-f(y) \|
\leq
L\| x-y \| ~~
\forall x,y \in \mathbb{R}^n
\label{eq:lips}
\end{equation}
\end{Definition}

The smallest  $L$  for which condition (\ref{eq:lips}) holds is called the Lipschitz constant  $L^*$, which provides an upper bound on how much the output  $f$  can change when the input varies from  $x$  to  $y$ . 
 {Denoting each element of $\mathbf\eta$ as $\eta^j, j\in \mathbb{I}^{M-1}_1$, 
we define
$
\eta^j = f^j(\boldsymbol{\theta}) = W^{l^j} \boldsymbol{s}^{l^j} + b^{l^j},
$
with $W^{l^j}\in \mathbb{R}^{1\times n_l}$ and $b^{l^j} \in \mathbb{R}$ the $j$-th row of $W^l$ and $b^l$.
We characterize the NN without the output layer using:}
\begin{equation}
\begin{split}
&\Sigma=
\begin{bmatrix}
W^0 & \cdots & 0 & 0\\
\vdots & \ddots & \vdots & \vdots\\
0 & \cdots & W^{l-1} & 0
\end{bmatrix}, 
V =
\begin{bmatrix}
0 & I_{n}
\end{bmatrix}
\label{eq:SigmaV}.
\end{split}
\end{equation}

\begin{Theorem}
If there exists \( {L^j}^2 > 0, T \in  \mathcal{D}_n\) such that:
\begin{equation}
	\begin{bmatrix}
\Sigma \\
V
\end{bmatrix}^T
F(T)
\begin{bmatrix}
\Sigma \\
V
\end{bmatrix}
+ 
\begin{bmatrix}
-{L^j}^2I & 0 & 0 \\
0 & 0 & 0 \\
0 & 0 & {W^{l^j}}^\top{W^{l^j}}
\end{bmatrix}
\preceq 0
\label{eq:LMI}
\end{equation}
then each $f^j$ is globally continuous with a Lipschitz constant \( L^j \geq L^{j*} \).
For a detailed proof, refer to \cite{fazlyab2019efficient}.
\end{Theorem}

Then the smallest value for each Lipschitz upper bound is obtained by solving the following semidefinite programs:
\begin{equation}
	\min_{{L^j}^2, T} {L^j}^2 \quad \text{s.t.} (\ref{eq:LMI})
\end{equation}
where \( {L^j}^2 \) and \( T \) are the decision variables and $j\in \mathbb{I}^{M-1}_1$.

 {
  {Limits $\bar\DDelta_h$ on the constraints relaxation~${\DDelta_h}_{n|k}$ set an upper bound $\bar\eta^j \in \mathbb{R}_+$ for each $\eta^j$. To force the NN to output a relaxation~$\eta^j\le\bar\eta^j$,} 
we impose the following constraint on \( L^j \)
\begin{equation}
    L^j \leq \bar L^j, \text{with}~\bar L^j=\frac{\bar\eta^j}{\bar\Delta_g+\bar\Delta_\x},
    \label{eq:Lj_bound}
\end{equation}
where $\bar\Delta_g$ specifies the maximum disturbance~$b_g$ 
and $\bar\Delta_\x = \max_{\substack{\x\in \mathrm{Proj}_\x{\Theta}, \u\in\mathcal{U}}} \norm{\x^\u-\x}$.
By enforcing the constraint~\eqref{eq:Lj_bound}, we guarantee that
when 
$\norm{
				\begin{bmatrix}
		{\Delta_g}_{n|k+1}
	\end{bmatrix}_{n\in \mathbb{I}_k^{k+M-1}}}\leq\bar\Delta_g$, then the inequality
\begin{equation*}\label{eq:eta_bound}
    \begin{split}
        \eta^j \leq 
        L^j
			\norm{
				\begin{bmatrix} 
				\begin{bmatrix}
		{\Delta_g}_{n|k+1}
	\end{bmatrix}_{n\in \mathbb{I}_k^{k+M-1}} \\
				\x_{k+1} - \x_{k}
			    \end{bmatrix}}
			\leq
            L^j (\bar\Delta_g+\bar\Delta_\x)
            \leq \bar\eta^j,
    \end{split}
\end{equation*}
holds, that is, the NN outputs a slack variable~$\DDelta_h\le\bar\DDelta_h$.
The inequality (\ref{eq:Lj_bound}) can be verified after the NN training phase or enforced during its training using the method proposed in \cite{pauli2021training}.
}

Denoting the output of the NN as \({\hat\DDelta_h}_{n|k}\), we assume the NN satisfies the following assumption.  
\begin{Assumption}\label{a:bound_est_error}
	The approximation error is uniformly bounded element-wise for the trained regressor, i.e.,  
	\begin{equation}
		|{\DDelta^\star}_{n|k} - {\hat\DDelta_h}_{n|k}|\leq \epsilon \mathbf{1}.
	\end{equation}
\end{Assumption}

We then formulate the following learning-based uncertainty-responsive SMPC control problem:
 {\begin{subequations}\label{eq:lunmpc}
		\begin{align}
		\min_{\substack{\bu}}& \sum_{n=k}^{k+N-1}
		q_\mathbf{r}(\xb,\ub)+p_\mathbf{r}(\xb[k+N])\nonumber\\
		\text{s.t.}\ &(\ref{eq:nmpcState}), (\ref{eq:nmpcDynamics}), (\ref{eq:nmpcInequality_unknown}), (\ref{eq:nmpcTerminalSafe})& \\
		&h(\xb,\ub) \leq{} E({\hat\DDelta_h}_{n|k}+\epsilon \mathbf{1}), & \hspace{-3em}~~n\in \mathbb{I}_k^{k+M-1} \\
		&(\xb. {\hat\DDelta_h}_{n|k}) \in {{\mathcal{Z}}}^\mathrm{s}_\mathbf{r},&\hspace{-3em}n\in \mathbb{I}_{k+N}^{k+M-1} 
		\end{align}
	\end{subequations}}

\subsection{Relaxed consistency condition and recursive feasibility}

We redefine safety as follows:
 {\begin{Definition}[Safety]\label{def:safe_2}
	A controller $\u_k = \kappa(\x_k)$ is safe for the system (\ref{eq:sys}) in a set $\mathcal{S}\subseteq\mathbb{R}^{n_x}$ if $\forall \, \x\in\mathcal{S}$, the control inputs $\mathbf{U}=\{\kappa(\x_0),...,\kappa(\x_\infty)\}$ and the corresponding state trajectories $\mathbf{X}=\{\x_0,\x_1,...,\x_\infty\}$ are such that $h(\x_k,\u_k)\leq{}E\bar\DDelta_h$ and~$g(\x_k,\u_k,{b_g}_k)\leq{}0$, $\forall \, k \geq 0$.
\end{Definition}}
 {
We then introduce the following assumption, which serves as a relaxed form of Assumption \ref{a:unknown_constraints}:
\begin{Assumption}\label{a:unknown_constraints_relaxed}
		The a-priori unknown constraint function satisfies
        \begin{align}
            g(\xb,\ub,{b_g}_{n|k+1}) &= g(\xb,\ub,\bg)+{\Delta_g}_{n|k+1},
        \end{align}
        where \begin{align}\label{eq:relax_g}\norm{
				\begin{bmatrix}
		{\Delta_g}_{n|k+1}
	\end{bmatrix}_{n\in \mathbb{I}_k^{k+M-1}}}\leq{\bar\Delta_g}_{k+1},\end{align}
        with
            ${\bar\Delta_g}_{k+1} = \min_j \frac{\bar\eta^j-\epsilon}{\bar L^j}-\norm{\x_{k+1}-\x_k}$.
\end{Assumption}
}

\begin{Theorem}\label{theorem:2} Suppose that Assumptions \ref{a:cont}-\ref{a:terminal}, and \ref{a:safe}-\ref{a:unknown_constraints_relaxed} hold, and that Problem~\eqref{eq:nmpc} is feasible for the initial state $\x_k$. Then, system \eqref{eq:sys} in closed loop with the solution of~\eqref{eq:lunmpc} applied in receding horizon is safe at all times.

\begin{proof}
Feasibility at time $k$ ensures that there exist $\mathbf{U}_{k}=\{\ub[k],...,\ub[k+M-1]\}$, $\mathbf{X}_k=\{\xb[k],...,\xb[k+M]\}$ and ${{\tilde\DDelta}_h}_{k}=\{ {\DDelta_h}_{k|k},...,{\DDelta_h}_{k+M-1|k}\}
$ that satisfy $h(\xb,\ub)\leq{} E{\DDelta_h}_{n|k}$, and $g_{n|k}(\xb,\ub)\leq{}0$
for all $n\in\mathbb{I}_k^{k+M-1}$. At time $k+1$, we consider two cases: 
\begin{enumerate}
    \item Assumption \ref{a:unknown_constraints} holds.
	\item Assumption \ref{a:unknown_constraints} does not but Assumption \ref{a:unknown_constraints_relaxed} holds.
\end{enumerate}

 {\textit{Case 1:} We verify that the trajectories
$\mathbf{X}^s_{k+1} = \{\xb[k+N+1],...,\xb[k+M],f(\xb[k+M],\u_\mathrm{safe}) \}$, $\mathbf{U}^s_{k+1} = \{\ub[k+N+1],...,\ub[k+M-1],\u_\mathrm{safe} \}$, and ${\tilde{\DDelta}^s_h}_{k+1}=\{ M_h{\DDelta_h}_{k+N+1|k},...,M_h{\DDelta_h}_{k+M-1|k}\}
$
satisfy $h(\xb,\ub)\leq EM_h{\DDelta_h}_{n|k}$, and $g(\xb,\ub,{b_g}_{n|k+1})\leq{}\gb[n][k](\xb,\ub)\leq{}0,\ \forall n\in\mathbb{I}_{k+N}^{k+M-1}.$
Since $\xb[k+M]\in\mathcal{X}_\mathrm{safe}$, Assumption~\ref{a:safe} ensures $h
(\x_{k+M|k+1},\u_\mathrm{safe})\leq{}0$, $\gb[k+M][k+1](\xb[k+M],\u_\mathrm{safe})\leq{}0$, and $\xb[k+M]^{\u_\mathrm{safe}}\in\mathcal{X}_\mathrm{safe}$. Thus, there exist control inputs $\u:=\ub[k+N]$ ensuring $(\xb[k+N]^\u, M_h{\DDelta_h}_{k+N|k})\in {{\mathcal{Z}}}^\mathrm{s}_\mathbf{r}$.
Recursive feasibility follows from Assumption~\ref{a:unknown_constraints}, ensuring that $\mathbf{U}_k$, $\mathbf{X}_k$, ${\tilde{\DDelta}_h}_k$ and their prolongation to infinite time satisfy constraints $h$ and $g$.}

 {
\textit{Case 2:} 
Assumption \ref{a:unknown_constraints_relaxed} ensures that $\eta^j_{n|k+1}
			\leq
            L^j (\bar\Delta_g+\norm{\x_{k+1} - \x_{k}})
            \leq \bar\eta^j-\epsilon$,
i.e., ${\hat\DDelta_h}_{n|k+1} + \epsilon \mathbf{1}
	\leq {\bar{\DDelta}_h}$.
At the same time, 
Assumptions \ref{a:fea_set} and \ref{a:bound_est_error} ensure that ${\DDelta^\star_h}_{n|k+1}$ exists and satisfies ${\DDelta^\star_h}_{n|k+1} 
	\leq {\hat\DDelta_h}_{n|k+1} + \epsilon \mathbf{1}
	\leq {\bar{\DDelta}_h} $ such that (\ref{eq:slack_opt}) is feasible, and the corresponding solutions 
$\mathbf{U}^\star_{k+1}=\{\ub[k+1][k+1]^\star,...,\ub[k+M][k+1]^\star\}$ 
and $\mathbf{X}^\star_{k+1}= \{\xb[k+1][k+1]^\star,...,\xb[k+M+1][k+1]^\star \}$
satisfying $h_{n|k+1}(\x^\star_{n|k+1},\u^\star_{n|k+1})\leq E{\DDelta^\star_h}_{n|k+1} \leq E\bar\DDelta_h$,  
$g_{n|k+1}(\x^\star_{n|k+1},\u^\star_{n|k+1}) \leq 0$,
therefore guaranteeing the safety of the system by Definition~\ref{def:safe_2}.
}
\end{proof}
\end{Theorem}

\subsection{{Priority-driven soft-constrained MPC}}
 {Let \( E^i \), \( i \in \mathbb{I}^l_1 \), denote the \( i \)-th relaxation choice, ranked from \( 1 \) (highest priority) to \( l \) (lowest priority). For each \( E^i \), two NNs are trained:  
the first NN, $\text{NN}^1_{E^i}$, is trained with \( \boldsymbol \theta \in  \Theta \), as defined in (\ref{eq:NN}), while the second NN, $\text{NN}^2_{E^i}$, acts as an infeasibility indicator, predicting whether the problem is feasible under \( E^i \), trained with both \( \boldsymbol \theta \in \Theta \) and \( \boldsymbol \theta \notin \Theta \). Denoting the output of $\text{NN}^2_{E^i}$ as \( \mathcal{F}^i \), we define:}
 {\begin{equation}
    \mathcal{F}^i = 
    \begin{cases}
        0, & \text{if (\ref{eq:unmpc}) with \( E = E^i \) is feasible,} \\
        1, & \text{if (\ref{eq:unmpc}) with \( E = E^i \) is infeasible.}
    \end{cases}
\end{equation}}

We introduce Algorithm \ref{alg:1} for the proposed priority-driven soft-constrained MPC.
\begin{algorithm}[t]
    \caption{Priority-driven Soft-constrained MPC}
    \label{alg:1}
    \textbf{Initialize:} Set \( \x_0 \).\\
    \For{each time step \( k \)}{
        Obtain \( \x_k \), evaluate \( g_{n|k}(\xb, \ub, {b_g}_{n|k}) \).\\
        \uIf{(\ref{eq:assump_unrelaxed}) holds}{ Solve (\ref{eq:nmpc}). }
        \ElseIf{(\ref{eq:relax_g}) holds}{  
            Compute \( \mathcal{F}^{i} \) via \( \text{NN}^2_{E^i} \) for \( i \in \mathbb{I}^l_1 \).  
            \For{lowest-priority \( i \) with \( \mathcal{F}^{i} = 0 \)}{  
                Solve (\ref{eq:lunmpc}) with \( E = E^i \), \( \hat\DDelta_h \) from \( \text{NN}^1_{E^i} \).  
            }
        }
        \Else{ Report failure, exit. }
    }
\end{algorithm}

\section{Simulation Result}
In this section, we present an autonomous driving example focusing on collision avoidance with an pedestrian at a crosswalk. The simulations are performed on a laptop computer using MATLAB, with the CasADi software package \cite{andersson2019casadi} and the IPOPT solver \cite{wachter2006implementation}. 

\subsection{Vehicle Model and System Constraints} 
We model the vehicle system longitudinal dynamics by defining the system state as $\x=
\begin{bmatrix}
    p & v & a
\end{bmatrix}^T$ and the control input as $\u = a^{\text {req}}$ , where $p$ represents the longitudinal position, $v$ is the longitudinal velocity, $a$ is the acceleration, and $a^{\text {req}}$ is the requested acceleration.
The system dynamics are defined as 
$
A=\left[\begin{array}{ccc}
0 & 1 & 0 \\
0 & 0 & 1 \\
0 & 0 & -t_{\mathrm{acc}}
\end{array}\right], \quad B=\left[\begin{array}{c}
0 \\
0 \\
t_{\mathrm{acc}}
\end{array}\right]
$
with $t_{\mathrm{acc}}=1.8~\si{\second}$  model constant that defines acceleration dynamics. 
We assume that system (\ref{eq:sys}) is subject to the following a-priori known box constraints:
   \begin{gather*}
     0\leq v \leq \bar v,   
   	\underline{a} \leq a \leq \bar a, 
   	\underline{a} \leq a^\mathrm{req} \leq \bar a,
    \underline{j} \leq j \leq \bar j,
    \underline{j} \leq j^\mathrm{req} \leq \bar j,
   \end{gather*}
where $j=a_+-a$ and $j^\mathrm{req}=a^\mathrm{req}_+-a^\mathrm{req}$ represent the actual and required jerks.
And the a-priori unknown constraint is set up to limit the vehicle’s position to prevent collisions, formulated with \( C_g = \begin{bmatrix}
    1 & 0 & 0
\end{bmatrix} \), $D_g=0$, and $b_g=p^{\text{obs}}$ where $p^{\text{obs}}$ represents the position of an obstacle, which may vary based on real-time observations.
We express the safe set as all states where the vehicle has come to a full stop, formulated as $\mathcal{X}_\mathrm{safe}:=\{ \,  \x \, | \, \x = A\x+B\u,
	h(\x,\u) \leq 0, \, v_{k+M|k} = 0 \, \}.$
For MPC design, we use a sampling time of $t_s = 0.05 \si{\second}$ and prediction horizons $N = 20$ and $M = 100$.

\subsection{Date Collection and Training}
 {We define two relaxation modes. One with $E^1$ relaxing only $\underline j$ with $\delta_j$, and the other one with $E^2$ relaxing both $\underline j$ with $\delta_j$, and $\underline a$ with $\delta_a$. 
For each mode ${E^i}$, two NNs, $\text{NN}^1_{E^i}$ and $\text{NN}^2_{E^i}$, are trained.
The input for each NN, $\text{NN}^j_{E^i}$, is $\boldsymbol \theta = \begin{bmatrix}
    p^\mathrm{obs}-p & v & a 
\end{bmatrix}^\top$.
The output of $\text{NN}^1_{E^1}$ is $\eta=\delta_j$, while the output of $\text{NN}^1_{E^2}$ is $\eta = \begin{bmatrix}
    \delta_j & \delta_a
\end{bmatrix}^\top$.
The NNs are trained offline on datasets generated by solving (\ref{eq:slack_opt}) across various scenarios. The dataset is constructed by varying parameters over specific predefined ranges.
The velocity $v$ ranges from $0$ to $5.5 \si{\meter\per\second}$, and the acceleration a spans from $-3.5$ to $0.1 \si{\meter\per\second\squared}$. The relative distance $p - p^{\text{obs}}$ is sampled between $0.1$ and $5 \si{\meter}$, while the requested acceleration $a^{\text{req}}$ varies from $-3.7$ to $2.5 \si{\meter\per\second\squared}$. Each parameter is incremented in uniform steps of $0.1$, providing a detailed dataset that captures a wide range of possible conditions for the NN training.
}

\subsection{Simulation scenario}
\begin{figure}[t]
\SetFigLayout{2}{1}
\centering
\subfigure{\includegraphics[width=0.9\columnwidth]{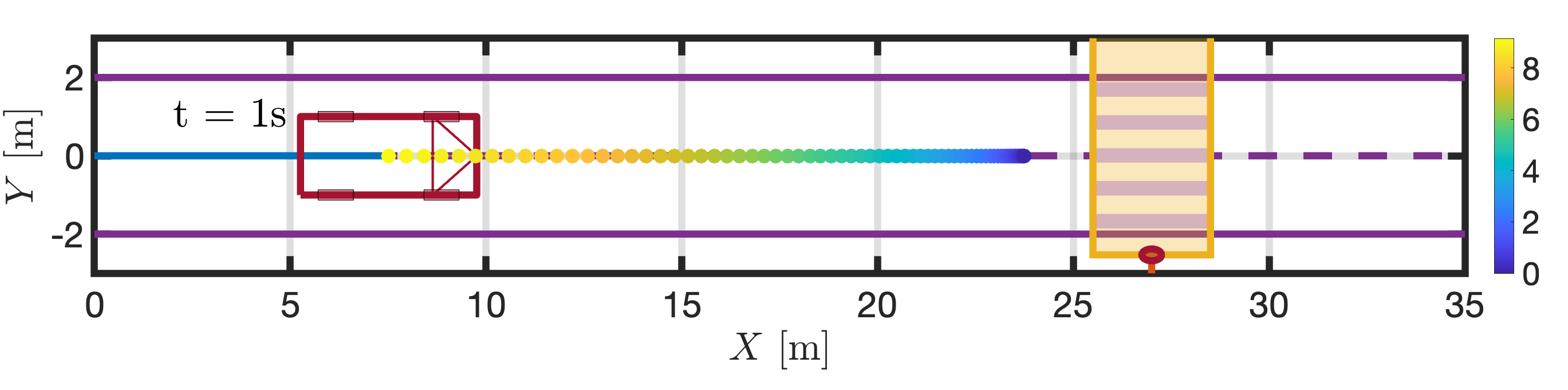}}
\hfill
\centering
\subfigure{\includegraphics[width=0.9\columnwidth]{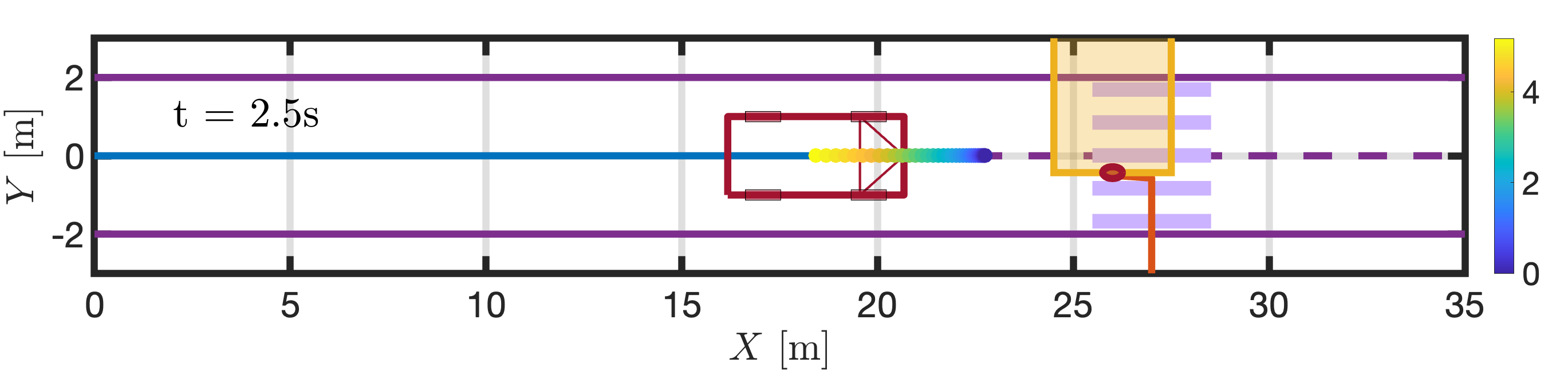}}
\caption{Two time instances of the scenario. Yellow shaded region denotes the predicted position of a pedestrian. Sensing the pedestrian, ego vehicle plans a trajectory (yellow/green/blue line indicating high/medium/low speed) within the sensing range.}
\label{fig:road}
\end{figure}

\begin{figure}[t] 
    \centering
    \includegraphics[width=0.9\columnwidth]{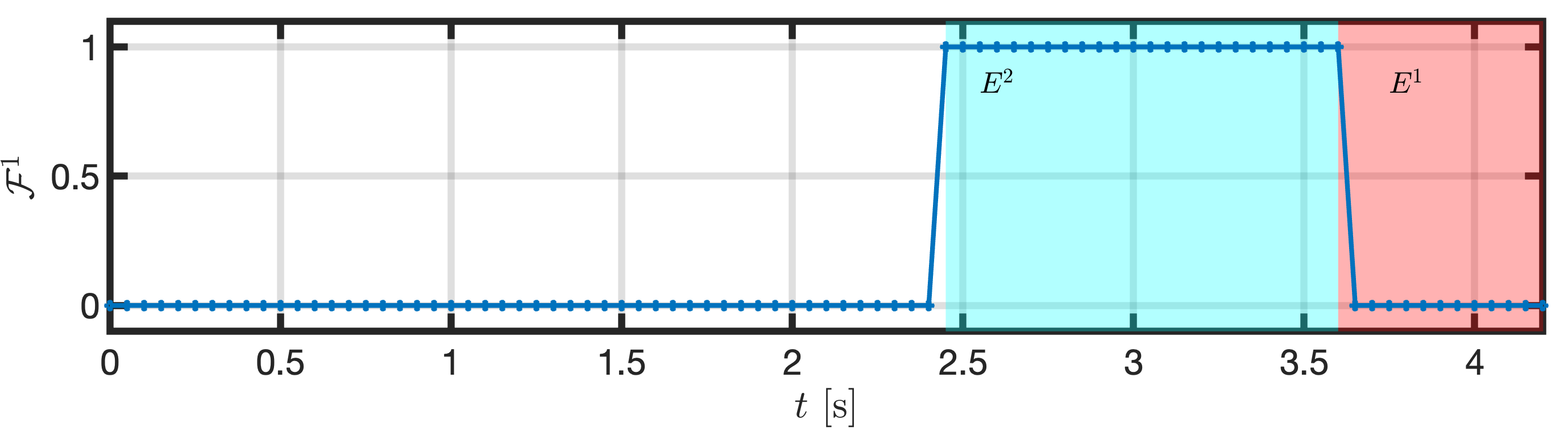}
    \caption{Infeasibility indicator $\mathcal{F}^1$ from $\text{NN}^1_{E^1}$.}
    \label{fig:FI}
\end{figure}

 {In this scenario, at \( t_1 = 2.5 \si{\second}\), a sudden change in pedestrian position is detected with $\Delta_g = 1\si{\meter}$, as shown in Fig. \ref{fig:road}. 
In Fig. \ref{fig:FI}, $\mathcal{F}^1 = 1$ at \( t_1 = 2.5 \si{\second}\) indicates that
adjusting only $\underline j$ is insufficient and will lead to an infeasible solution. 
To restore feasibility, $\text{NN}^1_{E^2}$ is deployed, updating both $\underline j$ and $\underline a$ to maintain feasibility. 
However, around \( t_2 = 3.6 \si{\second}\), $\mathcal{F}^1$ signals that feasibility can be maintained using applying $E^1$, prompting a switch to $\text{NN}^1_{E^1}$.
}

\begin{figure}[t]
\SetFigLayout{3}{1}
\centering
\subfigure{\includegraphics[width=0.9\columnwidth]{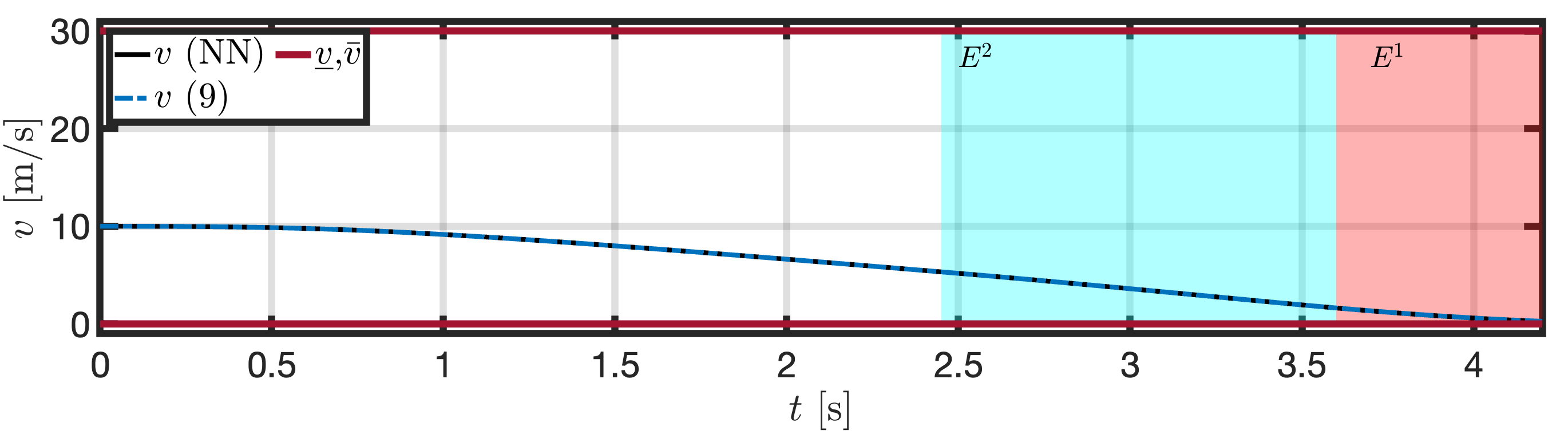}}
\hfill
\centering
\subfigure{\includegraphics[width=0.9\columnwidth]{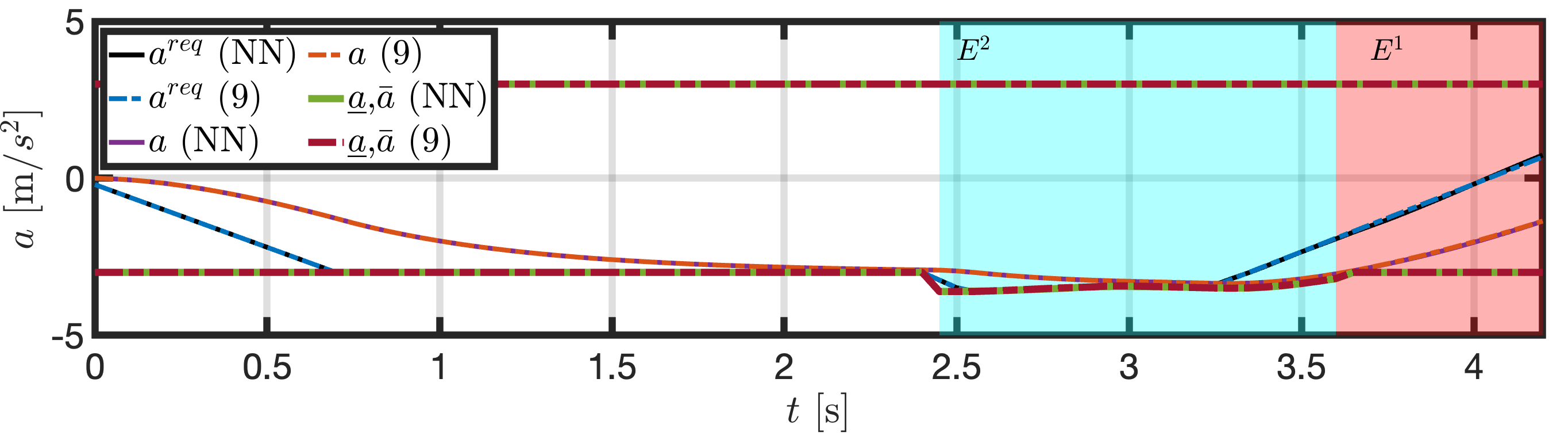}}
\hfill
\centering
\subfigure{\includegraphics[width=0.9\columnwidth]{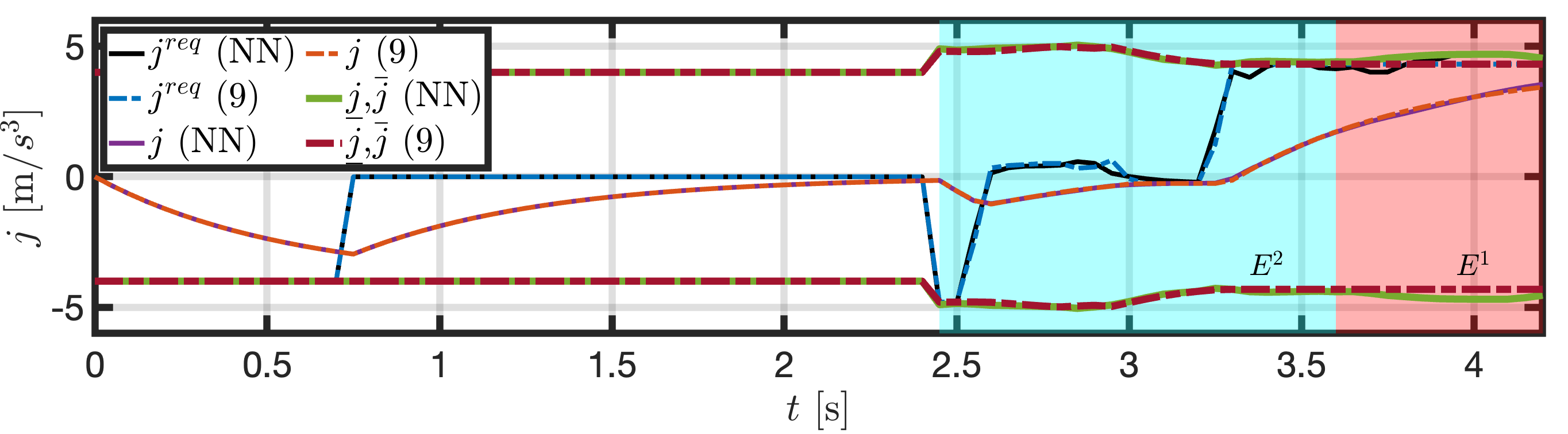}}
\caption{Closed-loop trajectories with NNs and (\ref{eq:unmpc}).}
\label{fig:Traj}
\end{figure}

\begin{figure}[t] 
    \centering
    \includegraphics[width=0.9\columnwidth]{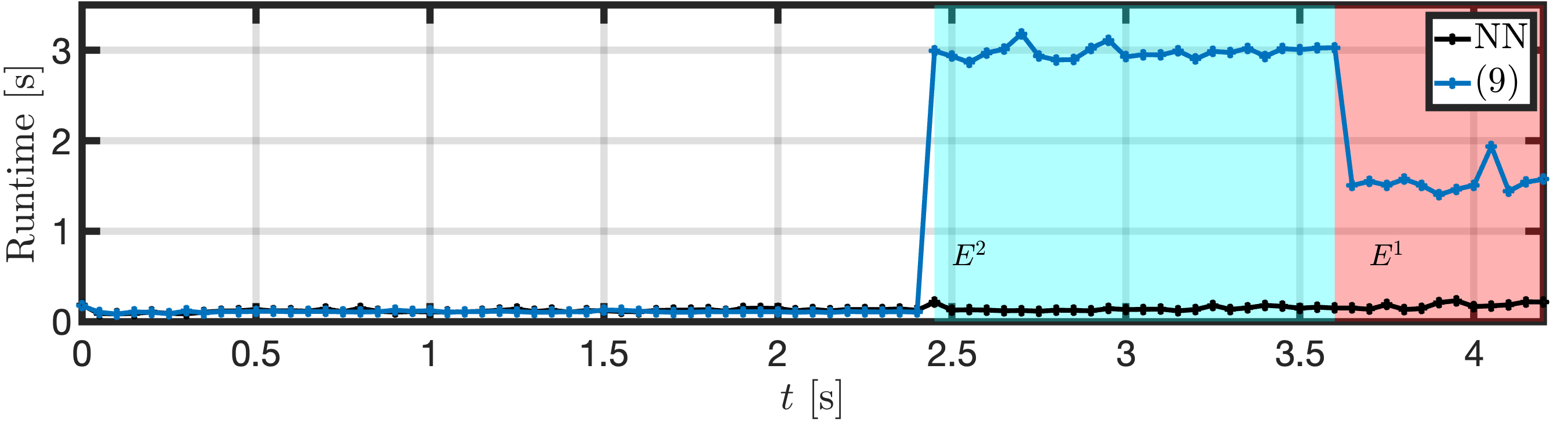}
    \caption{Computation time of applying  NN and (\ref{eq:unmpc})).}
    \label{fig:Runtime}
\end{figure}

 {We simulate the same scenario using the solution of (\ref{eq:unmpc}). The close-loop trajectories obtained by applying the NNs and (\ref{eq:unmpc}) are represented in Fig. \ref{fig:Traj}. At \( t_1 = 2.5 \si{\second}\), infeasibility is detected when solving (\ref{eq:unmpc}) with $E=E^1$, requiring a switch to $E=E^2$ until feasibility with $E=E^1$ is restored.
As seen in Fig. \ref{fig:Traj}, both methods result in the ego vehicle decelerating to a complete stop to avoid a collision with the pedestrian. However, the proposed method achieves this while reducing computational execution time, highlighting its potential for real-time applications, as shown in Fig. \ref{fig:Runtime}.}

\begin{figure}[!t]
\SetFigLayout{2}{1}
\centering
\subfigure{\includegraphics[width=0.9\columnwidth]{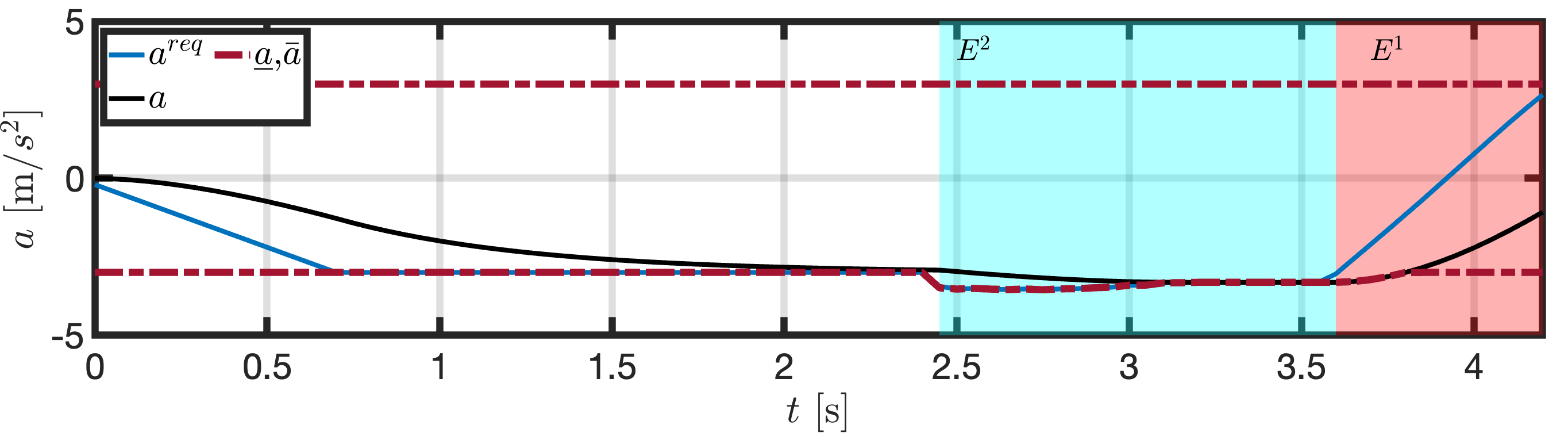}}
\hfill
\centering
\subfigure{\includegraphics[width=0.9\columnwidth]{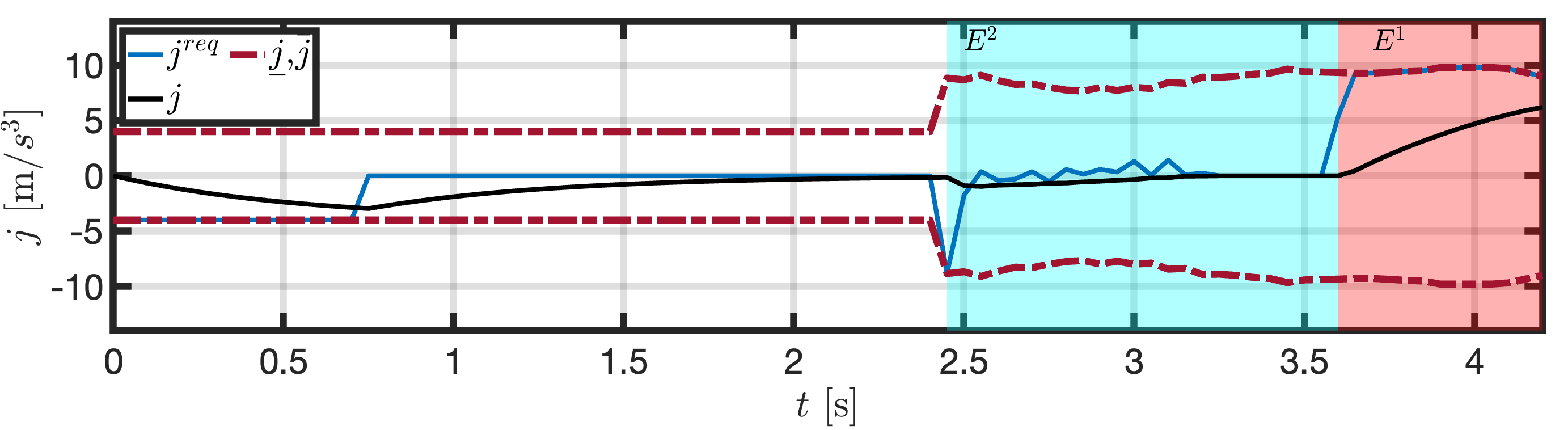}}
\caption{Closed-loop trajectories with softened MPC (design 1).}
\label{fig:soft_1}
\end{figure}

\begin{figure}[t]
\SetFigLayout{2}{1}
\centering
\subfigure{\includegraphics[width=0.9\columnwidth]{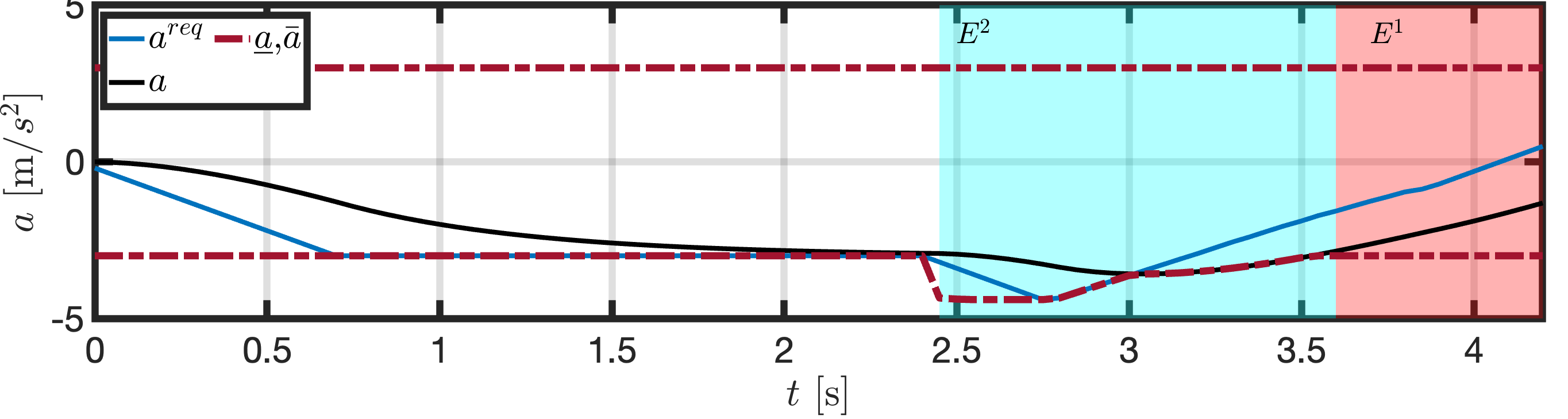}}
\hfill
\centering
\subfigure{\includegraphics[width=0.9\columnwidth]{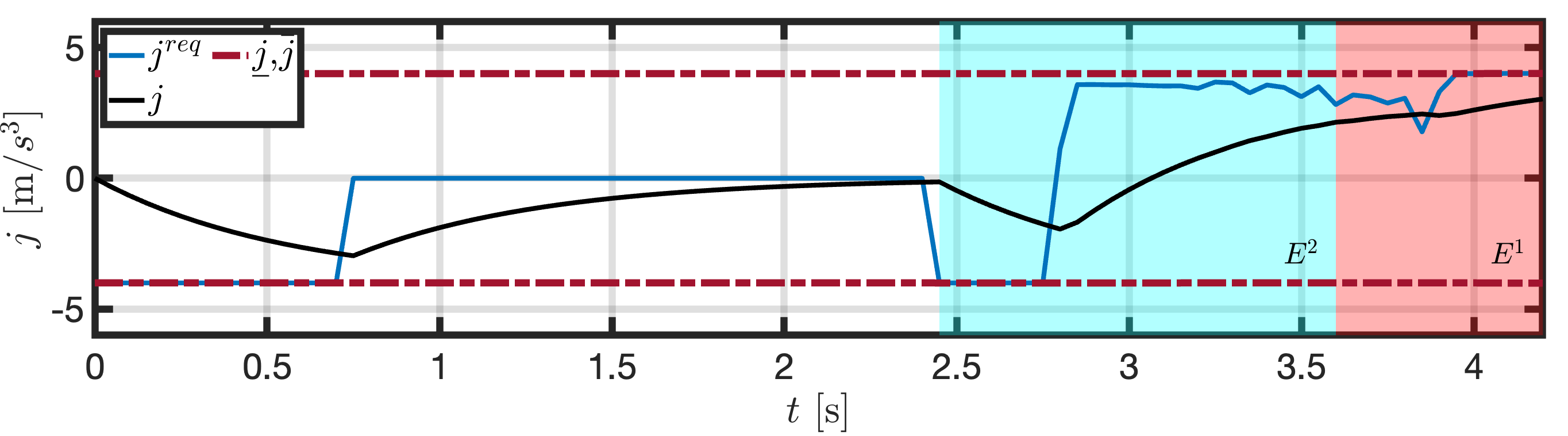}}
\caption{Closed-loop trajectories with softened MPC (design 2).}
\label{fig:soft_2}
\end{figure}

 {We further compare the results with those obtained by applying a softened MPC approach that introduces \(\delta_j\) and \(\delta_a\) as slack variables to relax constraints on \(\underline{j}\) and \(\underline{a}\), respectively. To analyze the effect of penalty weights $R_{\delta_j}$ and $R_{\delta_a}$ on slack variables, we consider two designs: design 1 with \(R_{\delta_j} = 10^1, R_{\delta_a} = 1.5 \times 10^3\) and design 2 with \(R_{\delta_j} = 1 \times 10^5, R_{\delta_a} = 10^1\).  
As shown in Fig. \ref{fig:soft_1} and Fig. \ref{fig:soft_2}, with appropriate tuning, softened MPC prioritizes relaxing \(\delta_j\) in the first case, while in the second case, it primarily relaxes acceleration constraints, resulting to a more conservative response.  
Adjusting the penalty weights enables some control over which constraint is relaxed more, producing results comparable to our approach (Fig. \ref{fig:soft_1}). However, strict prioritization cannot be guaranteed through tuning alone. Moreover, as the number of slack variables increases, identifying an appropriate tuning strategy that effectively enforces the desired prioritization becomes increasingly challenging \cite{kerrigan_soft_2000}.
}

\section{Conclusion}
In this paper, we proposed a priority-driven constraint softening for SMPC  that dynamically adjusts constraint relaxations in response to external disturbances. By integrating NNs to approximate the relationship between disturbances and constraint adaptations while assessing feasibility, our approach ensures safe and feasible control with reduced computational complexity.  
The simulation results of an autonomous driving scenario demonstrated the effectiveness of the proposed method in handling unexpected constraints while preserving safety.

\bibliographystyle{IEEEtran}

\begin{thebibliography}{10}
\providecommand{\url}[1]{#1}
\csname url@samestyle\endcsname
\providecommand{\newblock}{\relax}
\providecommand{\bibinfo}[2]{#2}
\providecommand{\BIBentrySTDinterwordspacing}{\spaceskip=0pt\relax}
\providecommand{\BIBentryALTinterwordstretchfactor}{4}
\providecommand{\BIBentryALTinterwordspacing}{\spaceskip=\fontdimen2\font plus
\BIBentryALTinterwordstretchfactor\fontdimen3\font minus \fontdimen4\font\relax}
\providecommand{\BIBforeignlanguage}[2]{{%
\expandafter\ifx\csname l@#1\endcsname\relax
\typeout{** WARNING: IEEEtran.bst: No hyphenation pattern has been}%
\typeout{** loaded for the language `#1'. Using the pattern for}%
\typeout{** the default language instead.}%
\else
\language=\csname l@#1\endcsname
\fi
#2}}
\providecommand{\BIBdecl}{\relax}
\BIBdecl

\bibitem{batkovic2022safe}
I.~Batkovic, M.~Ali, P.~Falcone, and M.~Zanon, ``Safe trajectory tracking in uncertain environments,'' \emph{IEEE Transactions on Automatic Control}, 2022.

\bibitem{batkovic2023experimental}
I.~Batkovic, A.~Gupta, M.~Zanon, and P.~Falcone, ``Experimental validation of safe mpc for autonomous driving in uncertain environments,'' \emph{IEEE Transactions on Control Systems Technology}, 2023.

\bibitem{liu2019recursive}
Z.~Liu and O.~Stursberg, ``Recursive feasibility and stability of mpc with time-varying and uncertain state constraints,'' in \emph{2019 18th European Control Conference (ECC)}.\hskip 1em plus 0.5em minus 0.4em\relax IEEE, 2019, pp. 1766--1771.

\bibitem{manrique2014mpc}
T.~Manrique, M.~Fiacchini, T.~Chambrion, and G.~Mill{\'e}rioux, ``Mpc tracking under time-varying polytopic constraints for real-time applications,'' in \emph{2014 European Control Conference (ECC)}.\hskip 1em plus 0.5em minus 0.4em\relax IEEE, 2014, pp. 1480--1485.

\bibitem{rakovic2021model}
S.~V. Rakovi{\'c}, S.~Zhang, H.~Sun, and Y.~Xia, ``Model predictive control for linear systems under relaxed constraints,'' \emph{IEEE Transactions on Automatic Control}, vol.~68, no.~1, pp. 369--376, 2021.

\bibitem{kerrigan_soft_2000}
E.~Kerrigan and J.~Maciejowski, ``Soft constraints and exact penalty functions in model predictive control,'' 09 2000.

\bibitem{borrelli_predictive_2017}
F.~Borrelli, A.~Bemporad, and M.~Morari, \emph{Predictive Control for Linear and Hybrid Systems}, 1st~ed.\hskip 1em plus 0.5em minus 0.4em\relax Cambridge University Press.

\bibitem{kerrigan_designing_2002}
E.~Kerrigan and J.~Maciejowski, ``Designing model predictive controllers with prioritised constraints and objectives,'' in \emph{Proceedings. {IEEE} International Symposium on Computer Aided Control System Design}.\hskip 1em plus 0.5em minus 0.4em\relax {IEEE}, pp. 33--38.

\bibitem{rawlings2017model}
J.~B. Rawlings, D.~Q. Mayne, M.~Diehl \emph{et~al.}, \emph{Model predictive control: theory, computation, and design}.\hskip 1em plus 0.5em minus 0.4em\relax Nob Hill Publishing Madison, WI, 2017, vol.~2.

\bibitem{fazlyab2019efficient}
M.~Fazlyab, A.~Robey, H.~Hassani, M.~Morari, and G.~Pappas, ``Efficient and accurate estimation of lipschitz constants for deep neural networks,'' \emph{Advances in neural information processing systems}, vol.~32, 2019.

\bibitem{pauli2021training}
P.~Pauli, A.~Koch, J.~Berberich, P.~Kohler, and F.~Allg{\"o}wer, ``Training robust neural networks using lipschitz bounds,'' \emph{IEEE Control Systems Letters}, vol.~6, pp. 121--126, 2021.

\bibitem{andersson2019casadi}
J.~A. Andersson, J.~Gillis, G.~Horn, J.~B. Rawlings, and M.~Diehl, ``Casadi: a software framework for nonlinear optimization and optimal control,'' \emph{Mathematical Programming Computation}, vol.~11, no.~1, pp. 1--36, 2019.

\bibitem{wachter2006implementation}
A.~W{\"a}chter and L.~T. Biegler, ``On the implementation of an interior-point filter line-search algorithm for large-scale nonlinear programming,'' \emph{Mathematical programming}, vol. 106, pp. 25--57, 2006.

\end{thebibliography}

\end{document}